\documentclass[twocolumn,nofootinbib,showpacs,preprintnumbers,superscriptaddress] {revtex4}

\usepackage{amssymb,amsmath,amsfonts,amsbsy,graphicx}
\usepackage{color}
\usepackage{enumerate}
\usepackage{ulem}

\newcommand{\dis}[1]{\begin{equation}\begin{split}#1\end{split}\end{equation}}
\newcommand{\be}{\begin{equation}}
\newcommand{\ee}{\end{equation}}
\def\bea{\begin{eqnarray}}
\def\eea{\end{eqnarray}}

\newcommand{\eq}[1]{Eq.~(\ref{#1})}

\newcommand{\bfrac}[2]{{\left(\frac{#1}{#2} \right)  }}
\newcommand{\VEV}[1]{\langle #1 \rangle}

\newcommand\tev{\,{\rm TeV}}
\newcommand\gev{\,{\rm GeV}}
\newcommand\mev{\,{\rm MeV}}

\newcommand\mpc{\,{\rm Mpc}}

\newcommand\kreh{k_{\rm reh}}

\newcommand{\Treh}{T_{\rm reh}}
\newcommand{\Tinf}{T_{\rm inf}}
\newcommand{\Tdom}{T_{\rm dom}}
\newcommand{\Trehcurv}{T_{\rm reh}}

\newcommand{\kdom}{k_{\rm dom}}

\begin{document}

\title{A new bound on the low reheating temperature with dark matter }

\author{Ki-Young Choi}
\email{kiyoungchoi@jnu.ac.kr}
 \affiliation{Institute for Universe and Elementary Particles and Department of Physics, Chonnam National University, 77 Yongbong-ro, Buk-gu, Gwangju, 61186, Republic of Korea}

\author{Tomo Takahashi}
\email{tomot@cc.saga-u.ac.jp}
\affiliation{Department of Physics, Saga University, Saga 840-8502, Japan}

\begin{abstract}

We investigate a new bound on the low reheating  temperature in a scenario where 
the Universe experiences early matter-domination before reheating after which  the standard big bang cosmology begins.
In many models of dark matter (DM), the small scale fluctuations of DM  grow during the early matter-domination era and seed the formation of the ultracompact minihalos (UCMHs).  
Using the constraints on the number of  UCMHs from gamma-ray observations, we find a lower bound on the reheating temperature between 
${\cal O}(10)~{\rm  MeV} -  {\cal O}(100)~{\rm  MeV}$ for WIMP dark matter
depending on the nature of DM.
A similar bound could be obtained for non-WIMP dark matter by observing UCMHs gravitationally such as pulsar timing, microlensing  and so on in some future observations.

\end{abstract}


\preprint{}

\maketitle

{\it Introduction}.---\quad
The early Universe is very well known for the temperature below  $1 \mev$. 
In other words,  the reheating temperature of the Universe must be  higher than this to be consistent with 
 current observations such as big bang nucleosynthesis, cosmic microwave background and large scale structure.

When the reheating temperature is lower than MeV,  the neutrinos are not thermalized fully and do not have the Fermi-Dirac distribution. This changes the proton-to-neutron ratio, then the  abundance of ${}^4$He, which sets the  limit on the reheating temperature as  $\Treh \gtrsim 0.7 \mev$ (or $\Treh \gtrsim 2.5\mev - 4 \mev$ in the case of  hadronic decays)~\cite{Kawasaki:1999na,Kawasaki:2000en}. With this low reheating temperature the oscillation of neutrinos can affect the thermalization too~\cite{Ichikawa:2005vw}.

By combining with cosmic microwave background and large scale structure data, the lower bound on the reheating temperature can be increased~\cite{Hannestad:2004px,Ichikawa:2006vm,DeBernardis:2008zz}. From the recent Planck data, the lower bound  was obtained as $\Treh \gtrsim 4.7 \mev$  when the neutrino masses are allowed to vary~\cite{deSalas:2015glj}.

The reheating of the Universe can happen in many situations  in the early Universe due to the decay of heavy non-relativistic particles. Some examples include the decay of inflaton field after inflation 
\cite{Kolb:1990vq,Linde:2005ht}
 or the decay of  heavy long-lived particles such as curvaton \cite{Enqvist:2001zp,Lyth:2001nq,Moroi:2001ct}, moduli or gravitino/axino~\cite{Coughlan:1983ci,Ellis:1986zt,deCarlos:1993wie,Choi:2008zq}. 
 The common feature of these reheating process is that an early matter-domination (eMD) by the decaying particles
precedes the reheating period and subsequent  radiation-domination.

When the reheating temperature is low enough, it is often the case that dark matter (DM) are already non-relativistic and decoupled from the relativistic thermal plasma. Since it is already decoupled and non-relativistic, the density contrast of DM can grow linearly in the scale factor within horizon during eMD which is much faster than that in  a usual radiation-domination epoch. 
DM density perturbations in this kind of scenarios have been discussed in \cite{Erickcek:2011us,Erickcek:2015jza,Choi:2015yma}. 
The enhanced perturbations then 
provide a higher probability to seed the small substructures  such as ultracompact minihalos (UCMHs), which are expected to survive to the present time~\cite{Ricotti:2009bs,Scott:2009tu}\footnote{
The formation of dense dark matter object in the RD era has also been investigated~\cite{Kolb:1994fi,Berezinsky:2010kq,Berezinsky:2013fxa}.
}. Therefore the determination of the present number of UCMHs can give clues to the early time of the Universe.

Up to now there is no convincing observation of small clumps of dark matter and this restricts the number of  UCMHs in the  Universe.
The bound on the fraction of UCMHs in the total matter, which we denote as $f$, 
was used to put constraints on the primordial power spectrum in the literatures. The strongest one comes from the  gamma-ray searches by the Fermi Large Area Telescope for weakly interacting massive particles (WIMPs), through the  annihilation of dark matters~\cite{Josan:2010vn,Bringmann:2011ut}. 
UCMHs can also be probed by purely gravitational methods such as 
the small distortions in the images of macrolensed quasar jets~\cite{Zackrisson:2012fa} with a constraint of  $f\lesssim 0.1$ for $k\sim 10^2\, \mpc^{-1}$,  astrometric microlensing~\cite{Li:2012qha} with 
$f\lesssim 0.01$ for $k\sim 10^3 - 10^4 \, \mpc^{-1}$ and the  pulsar timing~\cite{Clark:2015sha,Clark:2015tha}  with  $f\lesssim 0.01$ for $k  \sim 10^5 \, \mpc^{-1}$,
which might be observed in the future.

In this Letter, we suggest a new bound on the low reheating temperature, using the current and possible future constraints on the  abundance of UCMHs.

\bigskip
{\it Density perturbation during  early matter-domination}.---\quad
During the early matter-domination, 
the density perturbation of   dark matter grows linearly  with the scale factor, if they are already decoupled and non-relativistic.
$\delta_\chi \equiv  \delta \rho_\chi / \rho_\chi$ for a Fourier mode $k$ which enters the horizon during the eMD is given by
\dis{
\delta_\chi = -2\Phi_0 - \frac23\Phi_0\bfrac{k}{k_{\rm reh}}^2 , 
\label{swimp}
}
where $\Phi_0$ is the primordial gravitational potential and $k_{\rm reh}$ is that for the mode which enters the horizon at the time of reheating.
The wavenumber $k$ for a mode
 which enters the horizon during the eMD is related to the scale factor $a$ and the Hubble parameter $H$ as
\dis{
k = k_{\rm reh} \bfrac{a}{a_{\rm reh}}^{-1/2}
= k_{\rm reh} \bfrac{H}{H_{\rm reh}}^{1/3},
}
where 
$a_{\rm reh}$ ($a<a_{\rm reh}$) and $H_{\rm reh}$ are respectively the scale factor and the Hubble parameter 
at the time of reheating due to the decay of  non-relativistic heavy particle.
The scale of reheating $k_{\rm reh}$ has a relation to the reheating temperature as
\dis{
k_{\rm reh} =0.012\, {\rm pc}^{-1} \bfrac{\Treh}{\mev}\bfrac{10.75}{g_{*s}}^{1/3}\bfrac{g_*}{10.75}^{1/2},
}
where $g_*$ and $g_{*s}$ are effective  degrees of freedom of relativistic species and entropy, respectively.

When the scale enters before the beginning of the eMD, 
then the linear growth is limited to the epoch of eMD as
\dis{
\delta_\chi \simeq -\frac23 \Phi_0\bfrac{\kdom}{\kreh}^2 \quad {\rm for}\quad k > \kdom,\label{swimp2}
}
 where $\kdom$ denotes the scale which enters the horizon at the beginning of the eMD. 

For WIMPs, they could be still in kinetic equilibrium with relativistic plasma  for temperature around between MeV and GeV and the growth might be prevented even during early matter domination. However recent study~\cite{Choi:2015yma} shows that 
even in kinetic equilibrium, the subhorizon  isocurvature perturbation can be generated  during the eMD as
\cite{Choi:2015yma}
\dis{
\delta_\chi \simeq \frac54\Phi_0 \bfrac{k}{\kreh}^2,
\label{wimp}
} 
for the scales which enter the horizon during the eMD. 

However the density perturbations at small scales  are suppressed due to the
free streaming of dark matter.
For super-WIMP case where DM interacts superweakly such that they are already kinematically decoupled, the free-streaming scale 
can be calculated as~\cite{Kolb:1990vq} 
\dis{
k_{\rm fs}^{-1} =& \frac{1}{2\pi}
\int^{t_{\rm eq}}_{t_i} \frac{v}{a} dt 
\simeq
 \frac{1}{2\pi H_{\rm NR}a_{\rm NR}} \left[ 1+ \log\bfrac{a_{\rm dom}}{a_{\rm NR}} \right. \\
&\qquad \qquad \left.
 + 2\left(1-\sqrt{\frac{a_{\rm dom}}{a_{\rm reh}}}\right)
+\sqrt{\frac{a_{\rm dom}}{a_{\rm reh}}}\log\left(\frac{a_{\rm eq}}{a_{\rm reh}}  \right) \right]\\
\sim &\, 10^{-10} \bfrac{100\gev}{m_\chi}\bfrac{k_{\rm dom}}{k_{\rm reh}}^{1/2}\, \mpc,
\label{kfs_swimp}
}
where $t_{\rm eq}$ is the time at the radiation-matter equality and $t_i$ is  some initial time
much before eMD. The scale factor $a$ with subscript NR, dom, reh, and eq represent the time when DM becomes non-relativistic,  the beginning of eMD, the reheating epoch and the time of the radiation-matter equality,
 respectively. 
Here we assume that super-WIMP becomes non-relativistic before eMD begins\footnote{
When the super-WIMP becomes non-relativistic during the eMD, the mass dependance changes.
}. For WIMP, one can write it as~\cite{Erickcek:2015jza}
\dis{
k_{\rm fs}^{-1} =& \frac{1}{2\pi}\int_{t_{\rm kd}}^{t_{\rm eq}} \frac{v}{a} dt \simeq \frac{1}{2\pi}\sqrt{\frac{T_{\rm kd}}{m_\chi}}a(T_{\rm kd})
\int^{a_{\rm eq}}_{a(T_{\rm kd})} \frac{da}{a^3 H(a)}\\
\simeq &\, 7.7\times 10^{-8} \bfrac{100\gev}{m_\chi}^{1/2} \bfrac{\mev}{T_{\rm kd}}^{1/2}\, \mpc,
\label{kfs_wimp}
}
where $t_{\rm kd}$ is the time of the kinetic decoupling of WIMP dark matter with mass $m_\chi$, which 
is assumed to occur after  reheating.
Here $T_{\rm kd}$ is the temperature at $t_{\rm kd}$ and we put the scale factor at present as unity $a_0=1$.
DM fluctuations below this scale (i.e., $k>k_{\rm fs}$) are suppressed due to this free-streaming effect,
which can be taken into account by multiplying a factor
$ \exp\left( -k^2/2k_{\rm fs}^2 \right)$ to the transfer function of $\delta_\chi$.

\begin{figure}[!t]
\begin{center}
 \includegraphics[width=0.4\textwidth]{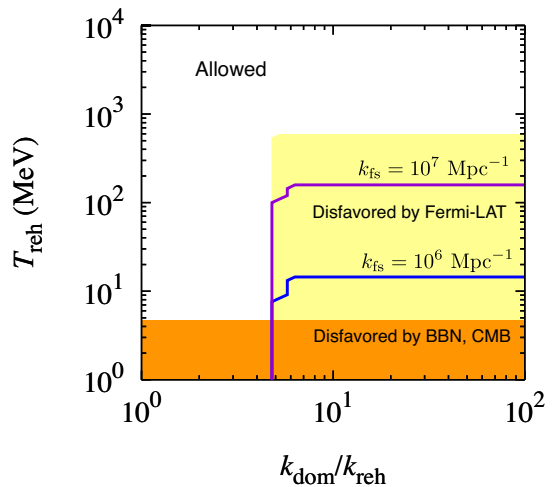}
\end{center}
\caption{Constraints on $T_{\rm reh}$ and $k_{\rm dom} / k_{\rm reh}$ for WIMP DM case.
The yellow region is disfavored from Fermi-LAT observation for the case with  $k_{\rm fs} > 5\times 10^7 \mpc^{-1}$  
where the free-streaming effect is negligible on the scale probed by the observations.
Cases with  $ k_{\rm fs} =10^6$ and $10^7\, \mpc^{-1}$ are also shown with purple and blue lines, respectively.
The orange regions is disfavoured by BBN and CMB observations.}
\label{fig:lowTreh}
\end{figure}

\begin{figure}[!t]
\begin{center}
 \includegraphics[width=0.4\textwidth]{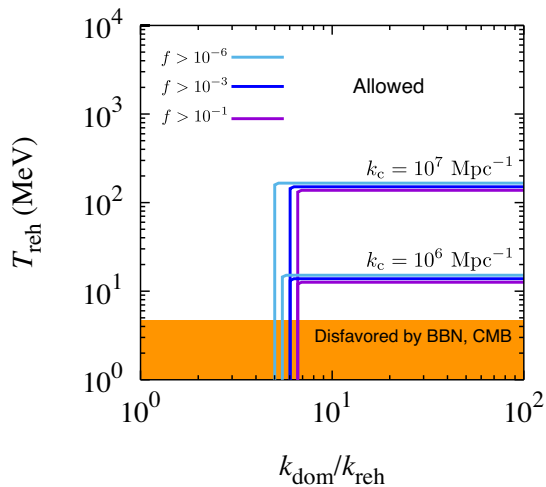}
\end{center}
\caption{
Projected bound on $T_{\rm reh}$ and $k_{\rm dom} / k_{\rm reh}$ given that $f$ is constrained as $f> 10^{-6}, 10^{-3}, 10^{-1}$  at the scale $k_c = 10^6$ and $10^7\, \mpc^{-1}$, 
which may be obtained from future observations by gravitational methods such as pulsar timing and gravitational lensing. Hence this bound can also be applicable to non-WIMP DM.
The orange region is disfavoured by BBN and CMB observations.
}
\label{fig:lowTreh_2}
\end{figure}

\bigskip
{\it Bound on the reheating temperature from UCMH.}---\quad
The growth of dark matter density fluctuations during the early matter-domination enhances  
the formation of UCMHs  after the radiation-matter equality. A large number of UCMHs can produce various signatures that can be detectable
by gamma-ray and cosmic ray for WIMPs or by gravitational interactions in  astrophysical observations, from which one can constrain the fraction of UCMHs in 
the total matter 
\begin{equation}
f \equiv \frac{\Omega_{\rm UCMH}}{\Omega_m},
\end{equation}
where $\Omega_m$ and $\Omega_{\rm UCMH}$ are the mass density of UCMH and matter in units of the critical density of the Universe.

For WIMP dark matter, the Fermi-LAT can put bounds   
on $f$ for scales from $k \simeq10~\mpc^{-1}$ to  $k \simeq 10^7~\mpc^{-1}$ which reaches as lowest as $f > 4 \times 10^{-7}$ at $k \sim 10^3~\mpc^{-1}$ 
for the annihilation cross section of $\VEV{\sigma v} = 3\times 10^{-26} $ cm$^3$ s$^{-1}$ of WIMPs into $b\bar{b}$ pairs~\cite{Bringmann:2011ut}.

One could also probe the abundance of UCMHs with gravitational ways such as pulsar timing, microlensing, 
small-scale distortion of macrolensed images \cite{Zackrisson:2012fa,Li:2012qha,Clark:2015sha,Clark:2015tha}
which could constrain $f$ in the future as 
\dis{
f\gtrsim 0.1 - 0.01,
}
for the scale around  $k\sim 10^2 - 10^6 \, \mpc^{-1}$ \cite{Zackrisson:2012fa,Li:2012qha,Clark:2015sha,Clark:2015tha}. 
These scales enter horizon around the cosmic temperature below $T\simeq 100\, \mev$  in the standard big bang Universe. This bound can be applied  for any kind of dark matter forming UCMHs since the observations are gravitational.

Here we briefly describe the formalism to constrain the UCMH abundance and the reheating temperature.
For details, we refer the readers to \cite{Bringmann:2011ut}.
Observations can put bound on the fraction of UCMH mass in our galaxy, which can be given as 
\begin{equation}
f  = \beta (R) f_\chi \frac{z_{\rm eq} +1 }{z_{c}+ 1}, 
\label{f_beta}
\end{equation}
where $f_\chi = \Omega_\chi/\Omega_m$  with $\Omega_\chi$ being the density parameter of dark matter and $z_c$ is the redshift at which the structure formation starts and the growth of the mass is assumed to be halted.
The factor $(z_{\rm eq} + 1)/(z_{c} + 1)$ corresponds to 
the growth of the mass by the infall of dark matter inside UCMH-forming region.
$\beta (R)$ is the probability of forming UCMHs for the region of comoving size $R$, which can be given by 
\begin{equation}
\beta (R) = \frac{1}{\sqrt{2\pi \sigma^2_{\chi,H} (R)}} \int_{\delta_{\rm min}}^{\delta_{\rm max}} \exp \left[ -\frac{\delta_\chi^2}{2 \sigma^2_{\chi,H}(R)} \right].
\label{betaR}
\end{equation}
Here $\sigma^2_{\chi,H}$ is the DM mass variance at horizon entry, which is calculated as
\begin{equation}
\sigma^2_{\chi,H} (R) = \int_0^\infty W_{\rm top-hat} ^2 (kR) \mathcal{P}_\chi (k) \frac{dk}{k},
\end{equation}
with $W_{\rm top-hat}(x) = 3(\sin x - x \cos x) /x^3$ being the top hat window function. 
Here the matter power spectrum $ \mathcal{P}_\chi (k) $ has a relation to that for the curvature perturbation $\mathcal{P}_R (k)$  as~\cite{Bringmann:2011ut} 
\begin{equation}
 \mathcal{P}_\chi (k) = \theta^4 T_\chi (\theta)^2 \mathcal{P}_R (k), 
\end{equation}
with $\theta = k R/ \sqrt{3}$ and $T_\chi (\theta)$ is the  transfer function for DM.

The $\delta_{\rm min}$ ($\delta_{\rm max}$) is the minimal (maximal) $\delta$ of dark matter for the formation of the UCMHs eventually.
The effect of the growth of DM density fluctuations  is accommodated in  the  $\delta_{\rm min}$.
In the standard Universe, $\delta_{\rm min}\sim 10^{-3}$ and $\delta_{\rm max}\sim 0.3$.
 However due to the growth during the eMD, in our scenario $\delta_{\rm min}$ can be lower.
To determine this, we follow the method in \cite{Bringmann:2011ut} with the modification to the transfer function in accordance with the evolution of $\delta_\chi$ given 
in Eqs.~\eqref{swimp} and \eqref{wimp}. 
To be conservative, we require that the collapse happens before the redshift $z_c=1000$.

Since $\beta\sim \exp(-\delta_{\rm min})$  for small $\delta_{\rm min}$, and $\delta$ grows as $\delta \sim k^2$ during the eMD, $\beta$ (therefore $f$ in \eq{f_beta}) is highly sensitive to the scale $k$. This means that the formation of UCMHs happens efficiently for a certain scale. When this scale overlaps with the scales constrained by observations,  the production of UCMHs is easily constrained. That is the reason of the sharp boundary at 
$k_{\rm dom}/k_{\rm reh}\sim 5$ in Figs.~\ref{fig:lowTreh} and~\ref{fig:lowTreh_2}.

\bigskip
{\it Case for WIMP dark matter}.---\quad
For the reheating temperature around GeV or below, the usual WIMP of 100 GeV mass is already chemically decoupled but they continue to be in the kinetic equilibrium until MeV. In this case, even in the kinetic equilibrium, the large isocurvature perturbation can be generated as in \eq{wimp} and lead to the formation of UCMHs~\cite{Choi:2015yma}. However the free-streaming of WIMP also erases the enhanced density perturbation and the formation of UCMHs on the scales smaller than $k_{\rm fs}$.
From \eq{kfs_wimp}, the free-streaming scale of WIMP with mass $100\gev$ and $T_{\rm kd}=1\mev$ is  $k_{\rm fs} \simeq 1.3\times 10^7
~{\rm Mpc}^{-1}$.

In Fig.~\ref{fig:lowTreh}, we show the constraints on the reheating temperature 
from Fermi-LAT which is applicable to WIMP dark matter. 
The yellow region below the corresponding free-streaming scale is disfavored and gives the lower bound on the reheating temperature. 
The orange region is disfavoured by the BBN and CMB observations. For enough growth of density perturbation to be visible, $k_{\rm dom}/k_{\rm reh}\gtrsim 5$ is necessary.

\bigskip
{\it Case for super-WIMP dark matter}.---\quad
Super-weakly interacting massive particles are  already decoupled for the temperature of our interest $T \lesssim {\cal O}(1)\gev$. Its famous examples  include gravitino, axino, axion or right-handed sterile neutrino dark matter~\cite{Baer:2014eja}. During the eMD, its density perturbation grows as~\eq{swimp}.

Super-WIMP dark matter has too small annihilation cross section to give sizable signatures in the gamma-ray observations. Instead the future observations using gravitational methods 
such as pulsar timing, microlensing and so on can be used to constrain the abundance of UCMHs made of super-WIMPs. In  Fig.~\ref{fig:lowTreh_2},
we show the expected bound on the reheating temperature given that the bound on the fraction of UCMHs as $f> 10^{-6}, 10^{-3}, 10^{-1}$  at the scales $k_c = 10^6$ and $10^7\, \mpc^{-1}$ respectively. The lower right region  below the lines would be  disfavoured.

We can see easily that the bound mostly depends on the observational scales rather than the value of the lower bound on the fraction $f$.
This is due to the exponential dependence of the probability to form UCMHs in \eq{betaR} on $\delta_{\rm min}$, which is proportional to  $k^2$.

The primordial relic density of super-WIMP which have existed from before the  early matter-domination is diluted by the entropy production due to the decay of heavy particles and a
new component is produced during or after reheating. The final dark matter abundance is the sum of both contributions
\dis{
Y_{\chi} =  \frac{S_i}{S_f}  \times Y_{\chi, 1}+ Y_{\chi,2},
}
where $Y_\chi = n_\chi / s$ is the ratio of the number density of dark matter  to the entropy density  $s=\frac{2\pi^2}{45}g_{*s}T^3$. $ Y_{\chi, 1}$ is the abundance of dark matter produced before the early matter domination (for example, at reheating stage after inflation) and $Y_{\chi,2}$ is the new dark matter abundance produced after early matter-domination. Here $S_i/S_f$ is the suppression due to the entropy production and in the sudden decay approximation it is given by~\cite{Kolb:1990vq}
\dis{
 \frac{S_i}{S_f}  \simeq \frac{T_{\rm reh}}{\Tdom}.
}

For gravitino or KSVZ axino, the dominant contribution comes from the highest temperature after reheating, which is denoted as $\Tinf$, and thus 
\dis{
Y_{\chi}(T_0) =  \left[\alpha \frac{\Tinf}{\Tdom}+\alpha\right] \Trehcurv \simeq \alpha \frac{\Trehcurv}{\Tdom} \Tinf  ,
}
where  we used $Y = \alpha (T) T$ and ignored the logarithmic dependence of $\alpha(T)$ on $T$. 
For the last equality
we used $\Tinf \gg \Tdom$.
Therefore, in spite of the entropy suppression, the gravitinos produced before the eMD can be still dominant. 
For given gravitino and gaugino mass, the dark matter relic density requires 
\dis{
\Tinf \simeq 10^{15}\gev\bfrac{\Tdom}{10^3\gev}\bfrac{10\mev}{\Trehcurv},
}
with $\alpha\simeq  7.4\times 10^{-23} \gev^{-1} $ with gravitino mass of 100 GeV and gluino mass $1\tev$.

Note that, in this case there could be large scale isocurvature perturbation between dark matter (produced from inflaton) and radiation (produced from heavy decaying field)
when the decaying field is independent on the inflaton field~\cite{Moroi:2002rd,Lyth:2002my,Lyth:2003ip}.
In such case, the super-WIMP dark matter is strongly constrained
from CMB~\cite{Ade:2015lrj}  and would be ruled out.

\bigskip
{\it Case for non-thermal  dark matter}.---\quad
Dark matter can be produced  non-thermally from decaying particles during reheating  process.
The evolution of the density perturbation depends on the velocity of dark matter.
When the produced  dark matter particles are relativistic, their fluctuations do not grow and there is no constraint on the reheating temperature.
In the  case of degenerate mass, the density perturbation of non-relativistic dark matter may grow~\cite{Erickcek:2011us} and the constraint on the reheating temperature can be obtained.  
In this case, its lower bound is
similar to that of super-WIMP case in Fig.~\ref{fig:lowTreh}.
For dark matter of bosonic coherent motion such as the axion, they are already non-relativistic  and the constraint in our study is applied~\cite{Kim:2008hd}.

\bigskip
{\it Case for light dark matter}.---\quad
The light thermal dark matter with mass below MeV is still relativistic for temperature larger than MeV. 
Therefore there is no enhancement in their density perturbation during the early matter-domination and no constraint on the reheating temperature from UCMHs is obtained.

\bigskip
{\it Conclusion}.---\quad
The reheating process in the early Universe accompanies an early matter dominated era  when the decoupled non-relativistic dark matter can have enhancement in their density perturbation. 
The growth of density fluctuations helps to form the UCMHs that can be probed by astrophysical observations. Using the current and possible future bound on the fraction of UCMHs, 
we could constrain the epoch of the early matter domination. Especially the reheating temperature can be constrained to be larger than around  ${\cal O}(10)\mev - {\cal O}(100)\mev$  for broad models of dark matter.

\bigskip
{\it Acknowledgments}.
TT would like to thank Teruaki~Suyama for helpful discussions on pulsar timing constraints on ultracompact mininhalos.
 KYC acknowledge the hospitality at APCTP where part of this work was done. KYC is supported by the National Research Foundation of Korea (NRF) grant funded by  the Korean government (MSIP) (NRF-2016R1A2B4012302). 
TT is partially supported by JSPS KAKENHI Grant Number
15K05084  and MEXT KAKENHI Grant Number 15H05888.

\end{document}